\documentclass{optica-article}
\newcommand{\be}{ \begin{equation} }
\newcommand{\ee}{\end{equation}}
\newcommand{\ci}{\textbf{i}}
\newcommand{\Dfrac}[2]{\displaystyle {\frac{\partial #1 }{\partial #2 }}} 
\newcommand{\Mdfrac}[2]{\displaystyle {\frac{d #1 }{d #2 }}}

\newcommand{\RomanNumeralCaps}[1]
    {\MakeUppercase{\romannumeral #1}} 
\journal{ome}

\articletype{Research Article}

\usepackage{lineno}

\begin{document}

\title{Nonequilibrium polariton condensation in biannular optically induced traps}

\author{Bochin A.\,K. ,\authormark{1} Nalitov A.\,V.\authormark{1,2}}

\address{\authormark{1} ITMO university,  Kronverksky pr., 49, lit. A, St. Petersburg, Russian Federation\\
\authormark{2}University of Wolverhampton, Wulfruna Street, Wolverhampton WV1 1LY, United Kingdom \\}

\email{\authormark{*}andrey.bochin@metalab.ifmo.ru} 



\begin{abstract}
We report the mean field model of nonequilibrium polariton condensation in annular effective non-Hermitian potential traps, stemming from incoherent optically induced excitonic reservoirs of annular shape.
We solve the linearized extended Gross-Pitaevskii equation in the approximation of two delta-function effective shell potentials for complex spectra of trapped polariton modes and calculate corresponding condensation threshold optical pumping powers.
The exhaustive map of condensate quantum number transitions in the multi-dimensional space of trap parameters, including a cascade of topological charge increments, is drastically different from the single annular trap case in topology and the range of accessible condensate states.
\end{abstract}

\section{Introduction}
Exciton-polaritons as mixed light-matter quasi-particles, formed in optical microcavities in the regime of strong coupling of cavity photon and exciton modes, present a peculiar platform supporting interacting bosonic condensates and low-threshold bosonic lasing \cite{Microcavities2017}.
In particular, nonequilibrium bosonic polariton condensates in effective potential traps, generated with profiled nonresonant optical pumping, can be simultaneously confined and populated via stimulated scattering by incoherent excitonic reservoirs \cite{Askitopoulos2013}.
This system supports surprisingly rich physical phenomenology, including Berry phase stemming from exceptional \cite{Gao2015} and diabolical points \cite{Estrecho2016}, formation of vortex lattices \cite{Sitnik2022}, spin bifurcations \cite{Ohadi2015,delValle2019}, spin ordering \cite{Ohadi2017,Sigurdsson2017} and topological transitions \cite{Pieczarka2021} in trap lattices.

Unlike their equilibrium counterparts, such polariton condensates are not necessarily formed in the ground state of a potential trap if one of its excited states is strongly populated by the reservoir due to a better overlap with the latter \cite{Askitopoulos2015a,Sun2018}.
In annular shaped traps this leads to formation of persistent counter-rotating polariton currents due to condensation in high angular momentum states \cite{Dreismann2014b,Li2016}.
In the nonlinear regime, where polariton interactions play a significant role, such condensates can evolve into space-time ordered phases \cite{Nalitov2019} and persistent vortex states \cite{Li2015}, whose topological charge can be controlled optically \cite{Dall2014,Ma2020} or, as in the case of spinor rotating condensates \cite{Yulin2016}, with magnetic field \cite{Yulin2020}.
Overall, nonequilibrium optically trapped polariton condensates offer a versatile platform for generation of coherent light with optically controllable spatiotemporal parameters, including angular momentum and topology.

The model suitable for description of polariton condensation in annular traps depends on their size, shape, and particular applications.
Elliptic traps of small radii, where condensation occurs at the ground state or at low excited states, are well described with the parabolic effective potential model \cite{Askitopoulos2018}.
In the case of larger traps the step-like effective potential model is reproduces the condensate angular momentum increasing with the trap size via a cascade of transitions from confined modes to continuum and to reproduce the superlinear dependence of the condensation threshold on the trap radius \cite{Nalitov2019}.
Delta function shell potential model also results in a ladder of angular momentum transitions in trap size and, in addition, allows assessing dissipative coupling of adjacent traps via overlapping evanescent condensate wavefunction tails \cite{Cherotchenko2021}.
In both step-like and delta function potential models the azimuthal periodicity of condensate density is explained in terms of counter-rotating vortex degeneracy lifted by imperfections of the cavity itself or the pumping profile.

In this work we address polariton condensation in biannular traps, formed by two concentric narrow shell potentials.
Such optically induced traps demonstrated possibility of condensation in the first excited radial quantum number state \cite{Dreismann2014b}.
We apply the non-Hermitian mean field model based on the extended Gross-Pitaevskii equation to compute complex energy spectra of the system.
Investigating the spectrum dependence on the density of the excitonic reservoir forming the trap, we then calculate the lasing threshold and identify the condensate parameters at this threshold.

The paper is organized as follows.
The theoretical model based on matching the solutions of extended Gross-Pitaevskii equation is described in Section \ref{theory}.
Section \ref{numerics} describes numerical solutions of the model and presents phase diagrams of condensate parameters.
Finally, the results and their implications are discussed in Section \ref{conclusions}.
\section{Theoretical model} \label{theory}
The nonequilibrium polariton condensate is described with the two-dimensional Gross-Pitaevskii equation (see, for example, \cite{Wouters2007a})
\be \label{GPE}
\ci\hbar\Dfrac{\Psi(t,\mathbf{r})}{t} = \left[-\frac{\hbar^2}{2m_p}\left( {\partial^2 \over \partial x^2}  + {\partial^2 \over \partial y^2}\right)+\frac{\alpha+\ci \beta}{2}N(t,\mathbf{r})-\ci\frac{\hbar\Gamma}{2}\right]\Psi(t,\mathbf{r}).
\ee
Here $\Psi(t,\mathbf{r})$ and $N(\mathbf{r})$ are the condensate wave function and the reservoir density respectively, $m_p$ is the effective polariton mass, $\alpha$ and $\beta$ are the condensate-reservoir interaction parameters describing repulsion and stimulated scattering from the reservoir into the condensate respectively, and $\Gamma$ is the polariton decay rate.

Eq. \eqref{GPE} is coupled to the semiclassical rate equation on the reservoir density
\be \label{res}
\Mdfrac{N(t,\mathbf{r})}{t}=P(\mathbf{r})-(\beta |\Psi|^2+\gamma)N(t,\mathbf{r}).
\ee

Here $\gamma$ is the excitonic reservoir decay rate and $P(\mathbf{r})$ is the spatially non-uniform pumping power.

In the vicinity of the condensation threshold pumping the condensate density may be neglected in Eq.\eqref{res} on the reservoir density.
In the adiabatic approximation the reservoir density relaxation time is assumed to be short compared to the characteristic timescales of the condensate dynamics.
The density of the reservoir thus instantly adjusts to the slowly evolving condensate, which allows expressing the quasi-stationary reservoir density as $N(t,\mathbf{r})=P(\mathbf{r})/(\beta |\Psi|^2+\gamma)$.
Substituting this expression, linearized in weakly populated condensate density $|\Psi|^2\ll\gamma/\beta$, in Eq.\eqref{GPE} results in local anti-Hermitian dissipative nonlinear terms in the effective Hamiltonian.
However, we are primarily interested in stationary states, where both condensed and reservoir parts of the polariton system are stationary.
Reservoir density in this case inherits the spatial profile of the pump and may be approximated by a superposition of two concentric delta function rings:

\be \label{res1}
N(r)=c[\delta(r-r_{1})+\delta(r-r_{2})],
\ee

Given the circular symmetry of the system the variables are separated in Eq.\eqref{GPE} with substitution $\Psi(t,r,\varphi)=e^{-\ci E t/\hbar + \ci m\varphi}R(r)$, where $\varphi$ and $r$ are polar coordinates, $m$ is the angular momentum, $E$ is the complex energy, and the radial wavefunction $R$ obeys the stationary dimensionless equation
\be
\label{eq_delta_potential_01}
\rho^2\Mdfrac{^2R}{\rho^2}+\rho\Mdfrac{R}{\rho}-\left\{\rho^2\cdot Z(\rho)+m^2\right\} R=0.
\ee
Here $\rho=r\sqrt{2\Gamma m_p/\hbar}$ is the normalized radius and $Z(\rho)$, in the case of a biannular trap with the radial profile formed by two delta functions, given by
\be
Z(\rho)= \frac{(\varkappa+\ci)}{2}\eta\left[\delta(\rho-\rho_1)+\delta(\rho-\rho_2)\right]-\left(\frac{\ci}{2}+\epsilon\right),
\ee
where $\varkappa={\alpha}/{\beta}$, $\epsilon={E}/{(\hbar\Gamma)}$, and $\eta=c\beta/\Gamma$ is the normalized pumping power.
The radial wavefunction converging at $\rho=0$ and $\rho\rightarrow\infty$ may be piecewise defined in three regions:
\be \label{definition_R}
R(\rho)=\left\{
\begin{array}{c}
       R_I(\rho) = A J_{m}\left(\rho\sqrt{\epsilon+\frac{\ci}{2}}\right), \quad \rho \leq \rho_{1}\\
       R_{II}(\rho) = C J_{m}\left(\rho\sqrt{\epsilon+\frac{\ci}{2}}\right)+D Y_{m}\left(\rho\sqrt{\epsilon+\frac{\ci}{2}}\right),\quad \rho_{1} \leq \rho \leq \rho_{2}\\
       R_{III}(\rho) = F H_{m}\left(\rho\sqrt{\epsilon+\frac{\ci}{2}}\right), \quad \rho_{2} \leq \rho
\end{array}\right.
\ee

The wavefunction continuity requires that $R_{\text{\RomanNumeralCaps 1}}(\rho_{1})=R_{\RomanNumeralCaps 2}(\rho_{1})$ and $R_{\RomanNumeralCaps2}(\rho_{2})=R_{\RomanNumeralCaps 3}(\rho_{2})$.
The other couple of conditions on the coefficients may be obtained by integrating Eq.\eqref{eq_delta_potential_01} in infinitesimal vicinities of $\rho_1$ and $\rho_2$:
\begin{align}
\label{System 1}
    &\rho^2_{1}\left[\Mdfrac{R_{\RomanNumeralCaps 2}(\rho_{1})}{\rho}-\Mdfrac{R_{\RomanNumeralCaps 1}(\rho_{1})}{\rho}\right]-\frac{\varkappa+\ci}{2}\eta\rho_{1}^2R_{\RomanNumeralCaps 1}(\rho_{1})=0,\\
   &\rho^2_{2}\left[\Mdfrac{R_{\RomanNumeralCaps 3}(\rho_{2})}{\rho}-\Mdfrac{R_{\RomanNumeralCaps 2}(\rho_{2})}{\rho}\right]-\frac{\varkappa+\ci}{2}\eta\rho_{2}^2R_{\RomanNumeralCaps 2}(\rho_{2})=0.
\end{align}

The above conditions form a linear system of equations on the coefficients of the wavefunction. It results in the relation between the complex energy $\epsilon$ and the pumping power $\eta$:
\be
\label{Key_equation}
\det(M(\epsilon,\eta))=0,
\ee
where the matrix $M$ is explicitly given in Appendix A.

\section{Numerical approach} \label{numerics}

The resonance condition Eq. \eqref{Key_equation} was numerically solved using two different approaches.
In the first approach, complex discrete spectra of energies $\epsilon$ were computed for varying values of pumping power $\eta$.
This allowed qualitatively illustrating the spectrum evolution with increasing pumping power and specifying the particular state that first crosses the condensation threshold $\mathrm{Im}\{\epsilon\}=0$.
Alternatively, mode-specific critical pumping powers were computed for a wide range of quantum numbers by solving \eqref{Key_equation} for real-valued $\epsilon$ and $\eta$.
The condensation threshold and the condensate parameters were studied in depth in this second approach by identifying the minimal critical pumping among all modes.
The results of both approaches are presented separately in the following subsections.

\subsection{Complex spectrum computation}\label{SS_Spectra}

Here we illustrate the behavior of polariton complex spectra in biannular optically induced traps.
Polariton complex spectrum in the presence of varying complex potential is discrete and may be labelled with integer non-negative angular number $m$ and radial number $n$.
Graphically represented evolution of the spectrum with increasing pumping power illustrates the competition of modes and the mechanism of polariton nonequilibrium condensation.

\begin{figure}[h]
\center{\includegraphics[width=0.5\textwidth]{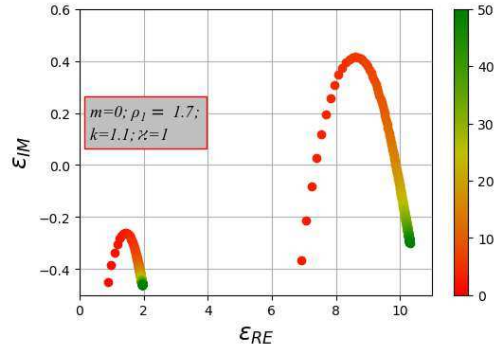}}
\caption{Numerically computed complex spectrum evolution with increasing pumping power indicated with colour.
The three visible series correspond to radial modes $n=0,1,2$.
Intersection of each line with the real axis $\epsilon_\textit{im}$ corresponds to a mode-specific critical pumping power and signifies possibility of polariton condensation at the mode.
Parameters: $m=0$, $\rho_{1}=1.7$, $\varkappa=1.1$, $k=1.0$.
}
\label{Spectrum2}
\end{figure}

Figure \eqref{Spectrum2} shows the evolution of the complex spectrum for $m=0$, $\rho_1 = 1.7$, $k =\rho_2/\rho_1= 1.1$, and $\varkappa = 1$ with increasing pumping power.
Note that each value of $n$ corresponds to a series of spectral points on the complex plane.
While the real part of the complex energy is monotonously increasing with pumping power, the imaginary part, corresponding to the growth rate ($\mathrm{Im}\{\epsilon\}>0$) or decay rate ($\mathrm{Im}\{\epsilon\}<0$), reaches its maximal value at a certain pumping power and decreases at further growing power.
If this maximal value is positive, the mode is characterized with a critical pumping power, where $\mathrm{Im}\{\epsilon\}=0$ and may form the condensate if this critical power is minimal among all modes governed by the quantum numbers $n$ and $m$.
These critical powers were numerically computed and are discussed in the following subsection.

For pumping powers above the condensation threshold multiple modes can simultaneously have positive imaginary parts of complex energies, which results in mode competition among such modes.
Although this regime is outside of the linearized model validity range, one may still expect the mode with the slowest effective decay rate (corrected for stimulated pumping by the reservoir) to form a stable condensate state if interactions weakly affect mode stability.
If, however, interactions play a significant role, multiple competing modes can dynamically mix, resulting in complex nonlinear behaviour.

\subsection{Critical pumping powers and quantum number maps} \label{SS_Thresholds}
In this part, we present the numerical results obtained by solving the equation \eqref{Key_equation} for real-valued $\epsilon$ and $\eta$ using the Levenberg–Marquardt algorithm.
We numerically computed the critical pumping values corresponding to various angular numbers $m$ in a range of trap radius values $\rho_1$ for fixed values radii ratio $k$ and $\varkappa = \alpha/\beta$.
Recovering the radial wavefunction $R(\rho)$ with extracted coefficients of the piecewise definition \eqref{definition_R}, we then attributed these states with radial quantum numbers $n=0,1,2,...$.
Finally, we computed the condensation threshold pumping power and the corresponding condensate quantum numbers $n$ and $m$ by minimizing critical pumping power values over all states in a region of the parameter space set by $\rho_1$ and $\varkappa$ for fixed values of $k$.

\begin{figure}[h]
\center{\includegraphics[width=7cm]{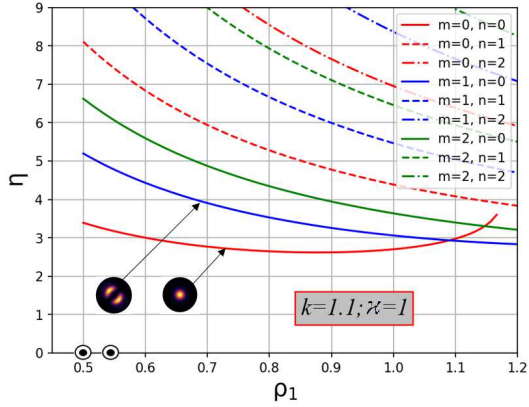}}
\caption{Polariton condensation critical pumping power $\eta$ dependence on the trap inner radius $\rho_{1}$ for angular momentum modes $m=0,1,...,5$ and $n = 0,1$.
The minimal value of $\eta$ among all modes at given trap size $\rho_1$ corresponds quantifies the condensation threshold and specifies the condensate quantum numbers ($n,m$).
An example of transition between two condensate modes ($0,0$) and $(0,1)$ with spatial density distributions shown in the insets, is indicated with the red dot.
Parameters: $k=1.1$, $\varkappa = 1$.}
\label{k=1.1}
\end{figure}

Figure \ref{k=1.1} shows the critical pumping dependence on the inner circle radius $\rho_1$ for fixed values of $k=1.1$ and $\varkappa = 1$.
The red dot indicates crossing of two graphs corresponding to $m=0$ and $m=1$ ($n=0$ for both) and illustrates switching between two condensate modes at $\rho_1\approx1.1$.
For smaller traps the condensation threshold corresponds to the lowest mode ($n=0$, $m=0$), while in larger traps the condensate forms at the mode ($n=0$, $m=1$).
The condensate density spatial distribution $|\Psi|^2$ of both modes is shown in the insets.
Further condensate quantum number switching is visible for larger traps at $\rho_1\approx1.6$.

\begin{figure}[]
\center{\includegraphics[width=8cm]{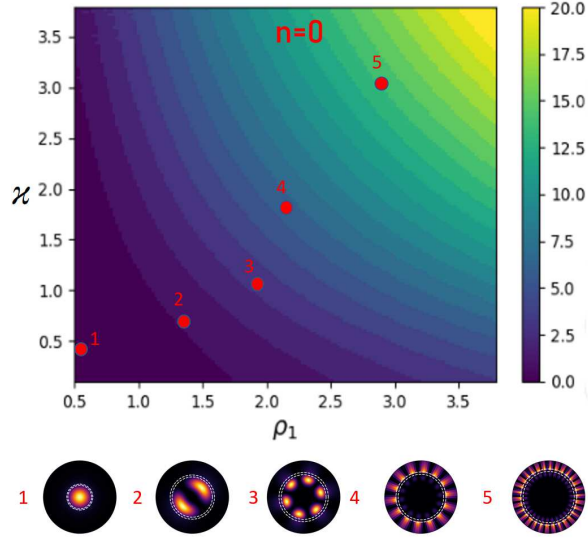}}
\caption{Angular quantum number distribution in the parameter space, formed by the trap inner radius $\rho_1$ and the interaction parameter ratio $\varkappa=\alpha/\beta$ in the case $k=\rho_2/\rho_1 = 1.1$.
The radial quantum number remains fixed ($n=0$) in the entire region of parameters.
Spatial condensate density distribution is shown with round insets for selected points on the plane.
}
\label{distr k=1.1}
\end{figure}

The case $k=1.1$, where the two circle radii are close, is qualitatively similar to the single annular trap, where the condensate only forms at the ground radial state $n=0$ \cite{Nalitov2019,Cherotchenko2021}.
This is illustrated in Figure \ref{distr k=1.1} showing the distribution of the condensate quantum numbers on the parameter plane  $(\rho_1,\varkappa)$ for $k=1.1$.
Similarly to the single annular trap case, switching cascade leads to the angular momentum $m$ increasing with the trap radius $\rho_1$ and $\varkappa$.
Insets illustrate the spatial density distribution of the corresponding condensate wavefunctions at selected parameter plane points.

\begin{figure}[h]
\center{\includegraphics[width=7cm]{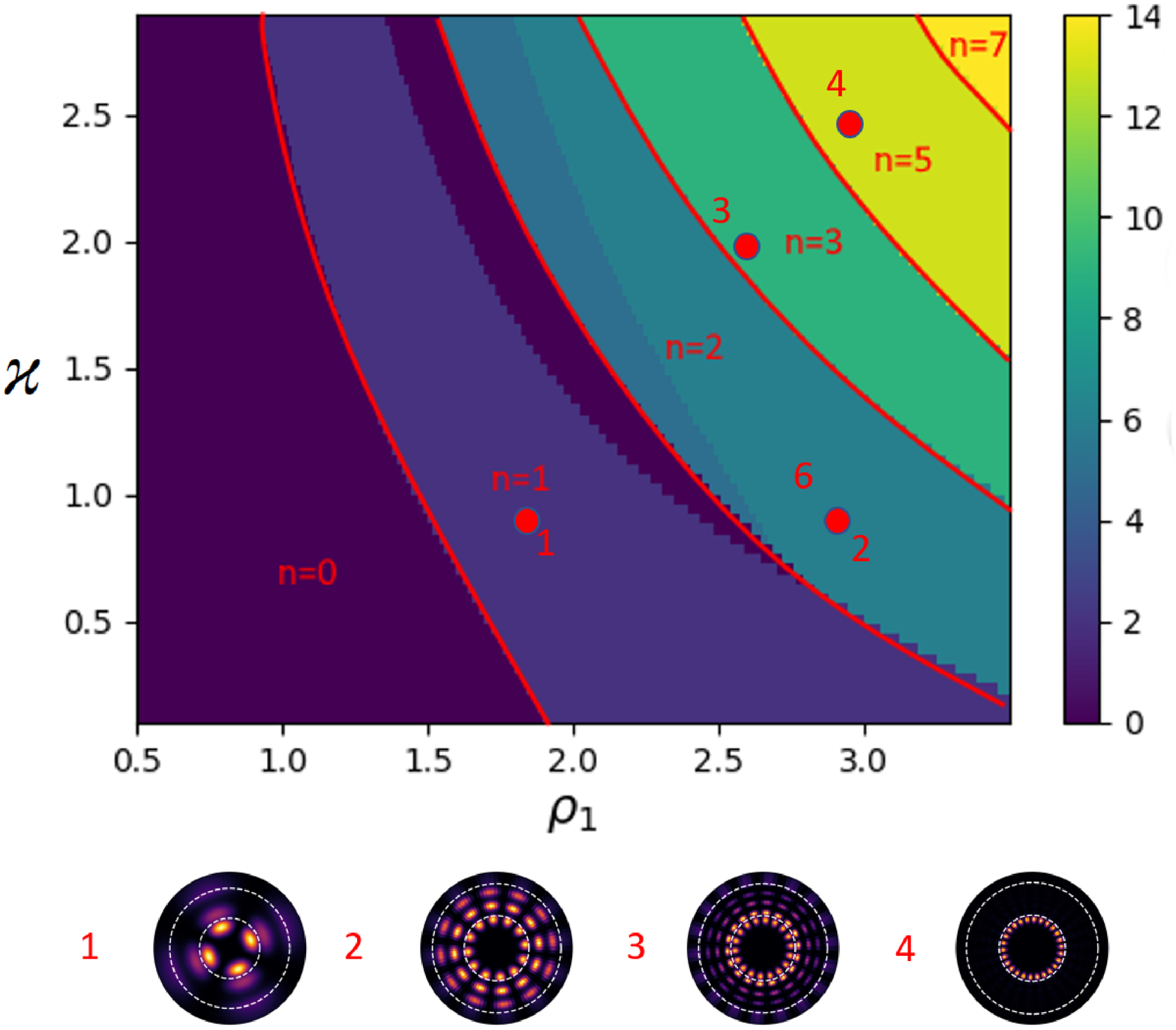}}
\caption{Distribution of quantum numbers $m$ (shown with colour) and $n$ in the parameter space, formed by the trap inner radius $\rho_1$ and the interaction parameter ratio $\varkappa=\alpha/\beta$ in the case of a higher ratio of radii $k=\rho_2/\rho_1 = 2$.
Spatial condensate density distribution is shown with round insets for selected points on the plane.}
\label{k=2}
\end{figure}

\begin{figure}[h]
\center{\includegraphics[width=7cm]{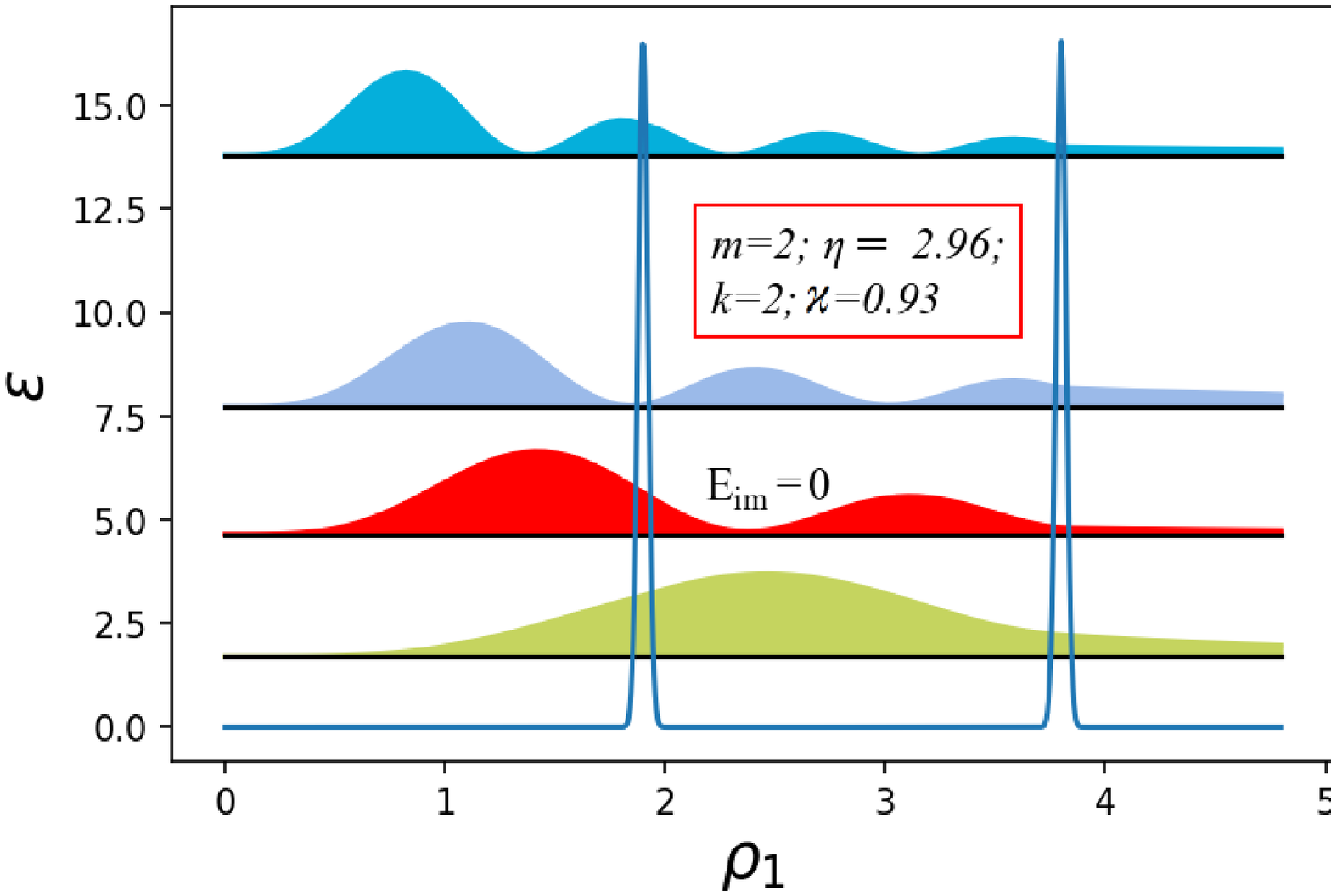}}
\caption{The form of wave functions near and at the critical point for a double delta-like potential. The graph shows the real parts of the energy levels sorted in ascending order.}
\label{Fig:levels}
\end{figure}

For higher values of $k$, the condensate quantum number distribution map significantly changes.
Figure \ref{k=2} shows the condensate quantum number distribution in the case $k=2$.
Most importantly, condensation at states with nonzero radial numbers $n\neq0$ is possible in certain domains of the parameter plane, in agreement with the experimental findings of Ref. \cite{Dreismann2014b}, which have not been reproduced in theory.
At the same time, monotonous growth of the azimuthal quantum number $m$ with $\rho_1$ and $\varkappa$ is replaced with a non-monotonous cascade of transitions, where both quantum numbers $n$ and $m$ change by varying increments.
One may also note that the distance between neighbouring transitions on the parameter plane generally increases.
Finally, in addition to continuous lines separating domains of condensate quantum numbers triple points lying on the edges of three domains emerge in drastic contrast to the single annular trap case.
Nontrivial behaviour of mode switching in this case if underlined by the fact that eigenstate wavefunction antinodes do not necessarily overlap with the two potential peaks.
This is illustrated in Fig. \ref{Fig:levels}, showing the real parts of the trapped polariton mode energies and corresponding probability density radial profiles.

\begin{figure}[h]
\center{\includegraphics[width=7cm]{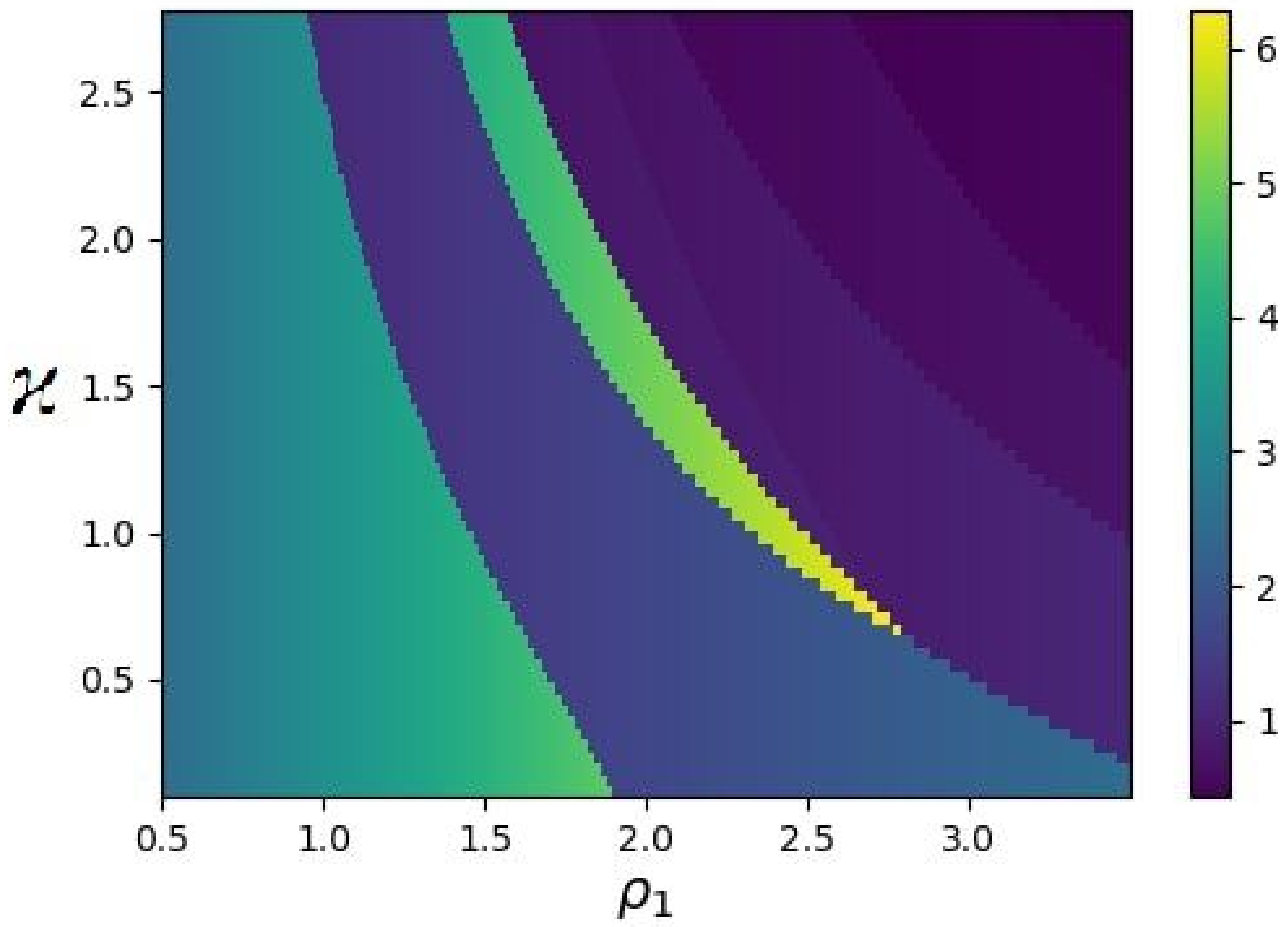}}
\caption{Numerically computed approximation for the distance between neighbouring azimuthal nodes of condensate density $d=\pi\rho_1/m$ in the same region of parameter space as in Fig. \ref{k=2} ($k=2$).}
\label{m and rho}
\end{figure}

Finally, the average distance between neighbouring azimuthal nodes of polariton condensate density spatial distribution was numerically estimated in the same region of parameter space.
Qualitatively, this parameter is approximately proportional to the ratio of the trap circumference $2\pi\rho_1$ and the number of azimuthal nodes $2m$.
The $\rho_1/m$ ratio distribution is presented in Fig \ref{m and rho}. 
Interestingly, although in each domain of quantum number distribution this value is linearly increasing with the trap radius $\rho_1$, at larger scales including many transitions the internodal distance is decreasing with the trap size in qualitative agreement with results of Refs. \cite{Sun2018,Dreismann2014b}.

\section{Discussion} \label{conclusions}

Polariton condensation was modelled in biannular optically induced complex potential traps within the framework of mean-field generalized Gross-Pitaevskii equation.
The model of biannular delta-function complex potential, emerging due to excitonic density generated by external nonresonant pumping, allowed quantitative in-depth analysis of polariton nonequilibrium condensation.
A drastic qualitative modification of polariton condensation picture in comparison to the single annular trap case was discovered.

This model allowed us to reproduce for the first time and to give qualitative interpretation to formation of polariton condensates in excited radial quantum number states of the trap, experimentally observed in Ref. \cite{Dreismann2014b}.
The developed numerical method allowed constructing phase diagrams of polariton condensate quantum numbers in arbitrary regions of the multidimensional space of optical trap parameters.
This paves the pave towards harnessing the radial degree of freedom of polariton laser emission via all-optical control of polariton condensate angular and radial quantum numbers.

One of the most intriguing predictions of the model is the modification of the condensate quantum number distribution map topology with increasing ratio of the biannular trap radii.
It supplements the layered cascade of incremental quantum number transitions with triple points in angular quantum number and quadrupule points if the radial quantum number is taken into account.
Topological characteristics of such points and the possibility of exceptional point emergence in this system remain unrevealed and may be studied within the developed model framework.

It should be noted that the discussed phenomena rely on both Hermitian and anti-Hermitian terms of the effective Hamiltonian, stemming from exciton-exciton repulsive interactions and bosonic stimulated particle exchange between the reservoir and the condensate respectively.
It is illustrated by qualitative difference of condensate mode switching patterns on the parameter $\varkappa$, quantifying the ratio of the two types of condensate-reservoir interaction.
Being an interacting non-Hermitian system, nonresonantly pumped nonequilibrium exciton-polariton condensates present an opportunity to vary this parameter by independently controlling exciton-exciton repulsion and stimulated scattering rates with the Hopfield coefficient on the one hand, and the energy dispersion anti-crossing shape on the other.

\section*{Appendix A}
The general view of the resulting matrix on the spectrum can be given as
\be
M=
\left(
\begin{array}{cccc}
M_{11}& M_{21} & M_{31} & 0 \\
0 & M_{22} & M_{32} & M_{42} \\
M_{13}& M_{23} & M_{33} & 0 \\
0 & M_{24} & M_{34} & M_{44} 
\end{array}\right).
\ee
\\First row $M_{11}= J_{m}\left(\rho_{1}\sqrt{\epsilon+\frac{\ci}{2}}\right), M_{21}=-J_{m}\left(\rho_{1}\sqrt{\epsilon+\frac{\ci}{2}}\right)$, $M_{31}=-Y_{m}\left(\rho_{1}\sqrt{\epsilon+\frac{\ci}{2}}\right)$
\\Second row $M_{22}= J_{m}\left(\rho_{2}\sqrt{\epsilon+\frac{\ci}{2}}\right) $, $M_{32}=Y_{m}\left(\rho_{2}\sqrt{\epsilon+\frac{\ci}{2}}\right)$, $M_{42}=-H_{m}\left(\rho_{2}\sqrt{\epsilon+\frac{\ci}{2}}\right)$
\\Third row $M_{13}= -J^{'}_{m}\left(\rho_{1}\sqrt{\epsilon+\frac{\ci}{2}}\right)-\frac{\varkappa+\ci}{2}\alpha_{1}J_{m}\left(\rho_{1}\sqrt{\epsilon+\frac{\ci}{2}}\right) $, $M_{23}=J^{'}_{m}\left(\rho_{1}\sqrt{\epsilon+\frac{\ci}{2}}\right)$, $M_{33}=Y^{'}_{m}\left(\rho_{1}\sqrt{\epsilon+\frac{\ci}{2}}\right)$
\\Fourth row
$M_{24}=-J^{'}_{m}\left(\rho_{2}\sqrt{\epsilon+\frac{\ci}{2}}\right)$, $M_{34}=-Y^{'}_{m}\left(\rho_{2}\sqrt{\epsilon+\frac{\ci}{2}}\right)$, $M_{44}=H^{'}_{m}\left(\rho_{2}\sqrt{\epsilon+\frac{\ci}{2}}\right)-\frac{\varkappa+\ci}{2}\alpha_{1}H_{m}\left(\rho_{2}\sqrt{\epsilon+\frac{\ci}{2}}\right)$


\begin{backmatter}
\bmsection{Funding}

This work was supported by the Russian Science Foundation under grant no.22-12-00144.

\bmsection{Acknowledgments}
The authors thank I.\,Chestnov for fruitful discussions.

\bmsection{Disclosures}
The authors declare no conflicts of interest.

\bmsection{Data Availability Statement}
The data underlying the results presented in this article are not currently publicly available, but may be obtained from the authors upon reasonable request.

\end{backmatter}








\bibliography{references} 

\end{document}